\pdfoutput=1
\documentclass[sigconf,nonacm]{acmart}
\AtBeginDocument{%
	}

\setcopyright{none}
\copyrightyear{2026}
\acmYear{2026}
\acmConference[AdKDD '26]{AdKDD 2026: Workshop on AI in Computational Advertising}{August 10, 2026}{Jeju, Republic of Korea}
\renewcommand\footnotetextcopyrightpermission[1]{}

\usepackage{algorithm}
\usepackage{algpseudocode}

\begin{document}
	
	
	\title{A Lightweight MPC Bidding Framework for Brand Auction Ads}
	\titlenote{This work has been accepted to ADKDD 2026.}
	
	\author{Yuanlong Chen}
	\authornote{Corresponding author}
	\email{ychenfd@gmail.com}
	\affiliation{%
		\institution{Bytedance Inc.}
		\city{San Jose}
		\state{Calinfornia}
		\country{USA}
	}
	
	\author{Bowen Zhu}
	\email{bowen.zhu@bytedance.com}
	\affiliation{%
		\institution{Bytedance Inc.}
		\city{San Jose}
		\state{Calinfornia}
		\country{USA}}
	
	\author{Bing Xia}
	\email{xiabing.07@bytedance.com}
	\affiliation{%
		\institution{Bytedance Inc.}
		\city{San Jose}
		\state{Calinfornia}
		\country{USA}}
	
	\author{Yichuan Wang}
	\email{yichuan.wang@bytedance.com}
	\affiliation{%
		\institution{Bytedance Inc.}
		\city{San Jose}
		\state{Calinfornia}
		\country{USA}}
	
	\renewcommand{\shortauthors}{Yuanlong Chen, Bowen Zhu, Bing Xia, Yichuan Wang}
	
	\begin{abstract}
		Brand advertising plays a critical role in building long-term consumer awareness and loyalty, making it a key objective for advertisers across digital platforms. Although real-time bidding has been extensively studied, there is limited literature on algorithms specifically tailored for brand auction ads that fully leverage their unique characteristics. In this paper, we propose a lightweight Model Predictive Control (MPC) framework designed for brand advertising campaigns, exploiting the inherent attributes of brand ads---such as stable user engagement patterns and fast feedback loops---to simplify modeling and improve efficiency. Our approach utilizes online isotonic regression to construct monotonic bid-to-spend and bid-to-conversion models directly from streaming data, eliminating the need for complex machine learning models. The algorithm operates fully online with low computational overhead, making it highly practical for real-world deployment. Simulation results demonstrate that our approach significantly improves spend efficiency and cost control compared to baseline strategies, providing a scalable and easily implementable solution for modern brand advertising platforms.
	\end{abstract}

	
	
	
	\maketitle
	
	\section{Introduction}
	Online advertising has become a cornerstone of modern commerce, enabling businesses to reach targeted audiences efficiently and at scale. Major technology companies such as Google, Meta, and TikTok rely heavily on digital advertising as a primary source of revenue, underscoring its critical role in the global economy. Advertisers engage in online advertising with the goal of achieving various marketing objectives, such as increasing brand awareness and engagement or boosting sales.
	
	To efficiently allocate limited advertising opportunities, auction mechanisms---such as second-price auction, first price auction, generalized second price auction---are widely employed to sell ad impressions. In modern digital advertising, these auctions are often conducted through a framework known as real-time bidding (RTB). RTB enables ad impressions to be auctioned dynamically as users visit websites or open mobile apps. For example, when a user performs a search on Google, scrolls through Instagram, or watches a video on TikTok, an ad request is triggered and sent to an ad exchange or the platform's internal bidding system. Multiple advertisers then rapidly evaluate the opportunity and submit bids based on the user's profile, context, and their campaign objectives. The platform runs an auction to determine the winning bid. The winning bidder earns the right to display their ad, typically paying according to the auction's pricing rule. Depending on the campaign setup, advertisers may be charged per impression (CPM) or per click (CPC). This entire decision-making process is completed within milliseconds to ensure a seamless user experience without noticeable delays.
	
	Performance ads and brand ads are the two main types of digital advertising, each playing a crucial yet distinct role in modern marketing strategies. Performance ads focus on driving immediate, measurable outcomes such as clicks, conversions, and sales. They rely heavily on real-time bidding, algorithmic optimization, and precise audience targeting to maximize return on ad spend (ROAS). However, performance ads often suffer from long feedback loops and data sparsity issues, as conversions may occur well after the ad impression and only a small fraction of impressions result in measurable actions. These challenges have been further intensified by increasing privacy regulations and restrictions on user-level tracking, such as GDPR, CCPA, and the phasing out of third-party cookies. Such regulations limit data availability and granularity, making it more difficult to accurately attribute conversions and optimize bidding strategies. In contrast, brand ads---including auction-based ads and guaranteed delivery contracts---are primarily designed to build long-term brand awareness, shape consumer perception, and foster loyalty. 
	
	Although real-time bidding strategies have been extensively studied in the literature, most existing methods focus on performance-driven campaigns and do not specifically address the unique characteristics of brand ads. Brand ads---such as video ads and awareness ads---exhibit distinct attributes, including fast feedback loops and abundant engagement data, which differentiate them from performance ads and open new opportunities for more responsive and data-efficient bidding strategies. However, tailored solutions for optimizing bidding in brand advertising remain largely unexplored. In this paper, we propose a lightweight Model Predictive Control (MPC) framework as an effective bidding strategy specifically designed for brand auction ads. This framework leverages the rich data availability and rapid feedback in brand campaigns to make informed, adaptive bidding decisions in real time.
	
	The rest of the paper is organized as follows. In Section 2, we review existing approaches to real-time bidding, highlighting their limitations in the context of brand advertising. In Section 3, we present our proposed lightweight MPC framework in detail. We begin by formulating the general bidding problem, followed by introducing methods to construct bid-to-\(X\) models---such as bid-to-spend and bid-to-conversions models---using a lightweight isotonic regression algorithm (Pooled Adjacent Violators Algorithm, PAVA). We then demonstrate how these models can be used to design effective bidding strategies for common bidding products, including maximum delivery and cost cap campaigns. In Section 4, we discuss how to evaluate the proposed algorithm through both offline simulations and online budget-split A/B testing experiments. In Section 5, we provide deeper insights into our MPC framework and then highlight the key challenges that arise when applying it to deep-funnel performance ads. Finally, in Section 6, we conclude the paper and outline future research directions.
	
	\section{Related Work}
	In the early stages of digital advertising, manual bidding was the predominant strategy, where advertisers set fixed bids based on their target objectives. While straightforward, this approach often led to inefficient budget utilization; if the bid or budget was not carefully calibrated, campaigns could either exhaust their budgets prematurely or underdeliver. To mitigate this issue, throttling-based pacing algorithms were introduced to control the rate of ad delivery over time, smoothing budget consumption and avoiding early campaign termination \cite{agarwal2014budget, xu2015smart} .
	
	As the advertising ecosystem evolved, automated bidding strategies became dominant. One research direction is to perform optimization at the individual campaign level by formulating the bidding problem as an online stochastic optimization problem, subject to constraints imposed by advertisers, such as budget limits and cost-per-action (CPA) targets \cite{balseiro2019learning, balseiro2017budget, castiglioni2022online, chen2025practical, conitzer2022multiplicative, fernandez2017optimal,  gao2022bidding, geyik2020impression, nuara2022online, wang2017display}. The existing literature on real-time bidding extensively studies these problems and proposes various algorithmic solutions, there are several popular approaches in industry: 1. Controller-based approaches, such as Proportional--Integral--Derivative (PID) controllers, which dynamically adjust bids to control the spend rate around a target value \cite{tashman2020dynamic, yang2019bid, zhang2016feedback}; 2. Dual-based online optimization methods, such as dual online gradient descent, which iteratively update dual variables to enforce budget and performance constraints \cite{balseiro2019learning, balseiro2022analysis, gao2022bidding}. 
	
	 Another direction adopts a more centralized perspective by formulating the problem as an online resource allocation task to optimize the overall objectives of the advertising platform~\cite{mehta2007adwords, aggarwal2019autobidding, chen2011real, abrams2008ad, grigas2017profit, feldman2010online}. In such approaches, the bidding strategy for each advertiser naturally emerges as a by-product of the platform-level optimization. 
	
	For branding ads, most existing work focuses on guaranteed delivery mechanisms under contractual agreements~\cite{chen2014dynamic, fang2019large, dai2024percentile, wang2022conflux, wu2021impression}, where the primary objective is to fulfill impression guarantees over a specified period. Other studies investigate the memory retention requirements of brand ads to optimize exposure frequency and timing to enhance long-term brand recall~\cite{maehara2018optimal}. In this work, we focus on auction-based branding ads and devise a lightweight, practical bidding strategy to address key objectives such as efficient budget pacing and cost-effective delivery in a dynamic auction environment.

	\section{Lightweight MPC Framework}
	We now present the technical details of the lightweight MPC framework for brand advertising optimization. Throughout this section, we assume that ad impressions are sold via standard second-price auction mechanisms. We begin by addressing the maximum delivery problem, where the campaign is constrained solely by its budget. This problem is formulated to maximize the campaign's welfare under the budget constraint, from which we derive the optimal bidding strategy. We then demonstrate how online isotonic regression can be leveraged to model the bid landscape and adaptively adjust bids within the MPC framework. Finally, we illustrate how this framework can be extended to handle more complex scenarios involving multiple constraints, such as cost caps.
	
	\subsection{Problem Formulation}
	We begin by formalizing the max-delivery problem. Consider a specific brand advertising campaign (e.g., an awareness ad or a video ad) that is charged on a \textbf{per-impression} basis. Let \( T \) denote the total predicted number of auction opportunities within a day. For the \( t \)-th auction, let \( r_t \) represent the \textit{welfare utility}---for example, \( r_t = 1 \) for awareness ads, or \( r_t \) equals the \textit{video play rate (VPR)} for video ads. Let \( c_t \) denote the cost associated with the \( t \)-th auction; in a second-price auction, this corresponds to the highest competing \textit{eCPM} bid. Define \( x_t \in \{0, 1\} \) as a binary decision variable indicating whether the campaign wins the \( t \)-th auction. 
	
	The welfare maximization problem under a budget constraint \( B \) can then be formulated as the following optimization problem:
	\begin{equation}  \label{eq:max_delivery}
		\begin{aligned}
			\max_{x_t \in \{0,1\}} \quad & \sum_{t=1}^{T} x_t \cdot r_t \\
			\text{s.t.} \quad & \sum_{t=1}^{T} x_t \cdot c_t \leq B.
		\end{aligned}
	\end{equation}
	
	Under certain regularity conditions (e.g., \( c_t \) and \( r_t \) are independently and identically distributed), it can be shown \cite{balseiro2019learning, gao2022bidding} that the optimal bid per impression at \(t\) is given by
	\begin{equation} \label{eq:optimal_bid}
	b^{*}_{t, imp} = \frac{r_t}{\lambda^*} = b^{*} \cdot r_t,
	\end{equation}
	where \( \lambda^* \) is the Lagrangian dual variable associated with the budget constraint in the primal-dual formulation, \(b^* = 1/\lambda^*\) is the optimal bid per conversion(we treat each impression in awareness ads as a ``conversion'' with rate 1) for this campaign. It can also be shown that this optimal bid level depletes the total budget exactly by the end of the campaign lifetime. This implies that, at a specific timestamp, the target spend rate per auction opportunity can be derived by dividing the remaining budget by the remaining auction opportunities. Traditional PID controllers and dual online gradient descent methods are designed based on this observation, adjusting bid or dual variable by examining the instantaneous gap between the actual spend rate and the target spend rate.
	
	In practice, instead of updating the bid for every auction request, it is more common to update in a \textit{batch} manner. That is, the bid per conversion or  the dual variable \( \lambda \) is updated at fixed time intervals of length \( \Delta t \), and the bid level remains unchanged within each pacing cycle. The pacing interval \( \Delta t \) is typically set between a few seconds and several minutes to balance the trade-off between update frequency and the variance of the actual spend rate within the pacing cycle.

	\subsection{MPC Framework}
	The general procedure of Model Predictive Control (MPC) operates as follows. At each time step, an optimization problem is solved to determine a sequence of control actions over a fixed time horizon. The first control input from this sequence is applied to the system; this approach is known as \textit{Receding Horizon Control (RHC)}. At the next time step, the process is repeated by solving a new optimization problem with the time horizon shifted one step forward, incorporating the latest system state and updated information. We mention here that one big advantage of MPC is its ability to handle constraints directly while requiring significantly less parameter tuning (e.g., controller gains, learning rates) compared to conventional control methods such as PID control and dual online grandient descent. 
	
	Specifically in our case, at the end of each pacing cycle \(\tau\), we adaptively reformulate our optimization problem (\textit{RHC}) as 
	
	\begin{equation} \label{eq: mpc_md_receding}
		\begin{aligned}
			\max_{x_t \in \{0,1\}} \quad & \sum_{t > \tau } x_t \cdot r_t \\
			\text{s.t.} \quad & \sum_{t > \tau } x_t \cdot c_t \leq B_{\tau}
		\end{aligned}
	\end{equation}
	where \(B_{\tau}\) is the remaining budget at \(\tau\).  Mathematically, this corresponds to the same optimization problem defined in \autoref{eq:max_delivery}. To determine the bid for the next pacing cycle of duration \( \Delta t \), we first compute the \textit{target spend rate}. The target spend for the upcoming interval \( \Delta t \) is set proportionally to the fraction of auction opportunities expected in that interval relative to the remaining campaign duration. Specifically, the target spend \(TS_{\tau}\) in \(\Delta t\) satisfies
	\begin{equation} \label{eq:target_spend}
	 TS_{\tau} = B_{\tau} \cdot \frac{N_{\tau, \Delta t}}{T-\tau},
	\end{equation}
	where \( N_{\tau, \Delta t} \) denotes the predicted number of auction requests in the next pacing cycle. 
	
	Suppose we have a bid landscape model 
	\[
	s = f(b),
	\]
	which characterizes the relationship between the bid level \( b \) (here corresponding to \( 1/\lambda \), without accounting for conversion rate) and the resulting spend rate \( s \). To achieve a desired target spend rate \( TS_{\tau} \) in the next pacing interval, the bid can be updated by applying the inverse of the bid landscape function:
	\[
	b_{\tau} = f^{-1}(TS_{\tau}).
	\]

	\subsection{Lightweight Bid-to-\(X\) Modeling}
	We now show how to construct such bid-to-\( X \) models using a lightweight \textit{isotonic regression} approach \cite{barlow1972statistical}. Here, \( X \) can represent various performance metrics over a fixed pacing interval, such as the amount of spend, the number of impressions, or the number of video plays.  We use spend as an example to demonstrate the idea, the cases of impression and video play can be handled in the same way. 
	
	 We leverage the most recent pacing data from this campaign. Recall that, within each bid update interval $\Delta t$, the bid per click $b_t$ remains unchanged. We may collect the most recent $N$ interval bid-spend pairs $\{b_k, s_k\}$, where $s_k$ represents the spend over the fixed interval $\Delta t$ at time $k$ (and thus can be considered as the spend rate). The bid-spend rate relationship for the next interval $\Delta t$ can then be learned from these $N$ data points, assuming that the number of auction opportunities does not change too much over small intervals of $\Delta t$. For a specific auction, if we bid higher, the spend should increase (or at least not decrease). Therefore, $f(b)$ should be a monotonically nondecreasing function.
	 
	  However, $\{b_k, s_k\}$ does not necessarily form a monotonic sequence, as the data points are collected from different time intervals. Isotonic regression is a type of regression analysis designed for situations where the target variable is expected to be non-decreasing (or non-increasing) with respect to an independent variable. It imposes an order constraint on the data, ensuring the resulting function remains monotonic. A key algorithm used for isotonic regression is the \textbf{Pool Adjacent Violators Algorithm (PAVA)} \cite{ayer1955empirical, busing2022monotone}. PAVA efficiently solves isotonic regression problems by iteratively merging adjacent data points that violate the monotonicity constraint. It is computationally lightweight, operating in linear time $O(n)$, making it well-suited even for large datasets. 
	  
	 Once the monotonic bid-spend sequence is obtained, the function \( f(b) \) can be constructed as a monotonic curve by performing linear interpolation between the available data points and extrapolation outside the observed range. The algorithms for constructing this monotonic mapping and computing the corresponding bid values are summarized in Algorithm \autoref{alg:pava} and Algorithm \autoref{alg:mpc_bid_update_pava}.
	  
		\begin{algorithm}[t]
			\caption{Pool Adjacent Violators Algorithm (PAVA)}
			\label{alg:pava}
			\begin{algorithmic}[1]
				\State \textbf{Input:} $\{(b_k, s_k)\}_{k=1}^{n}$: Bid-spend pairs sorted in ascending order by bid
				\State \textbf{Output:} $\{(b'_k, s'_k)\}_{k=1}^{m}$: Monotonic bid-spend pairs
				\State Initialize $s'_k \gets s_k$ and $w_k \gets 1$ for all $k$
				\State $i \gets 1$
				\While{$i < n$}
				\If{$s'_i > s'_{i+1}$} \Comment{Monotonicity violation}
				\State Merge:
				\[
				s'_i \gets \frac{w_i s'_i + w_{i+1} s'_{i+1}}{w_i + w_{i+1}}, \quad w_i \gets w_i + w_{i+1}
				\]
				\State Remove $s'_{i+1}$, $w_{i+1}$, and $b_{i+1}$
				\State $i \gets \max(1, i-1)$
				\Else
				\State $i \gets i + 1$
				\EndIf
				\EndWhile
				\State \Return $\{(b'_k, s'_k)\}$
			\end{algorithmic}
		\end{algorithm}
		

	\begin{algorithm}[t]
		\caption{MPC-based Max Delivery Bidding Algorith}
		\label{alg:mpc_bid_update_pava}
		\begin{algorithmic}[1]
			\State \textbf{Input:} 
			\quad$\{(b_k, s_k)\}_{k=1}^{n}$: Bid-spend pairs sorted in ascending order by bid, 
			$B_{\tau}$: remaining budget at cycle $\tau$, 
			$N_{\tau,\Delta t}$: predicted auction requests in next interval, 
			$T$: total number of auction opportunities
			\State \textbf{Output:} $b_{\tau}$: computed bid for interval $\Delta t$
			\vspace{0.5ex}
			\State \textbf{Compute target spend:}
			\[
			TS_{\tau} \;\gets\; B_{\tau}\,\times\,\frac{N_{\tau,\Delta t}}{\,T - \tau\,}
			\]
			\vspace{0.5ex}
			\State \textbf{Apply PAVA:}  
			\quad Use Algorithm~\ref{alg:pava} on $\{(b_k, s_k)\}$ to get monotonic $\{(b_k, s'_k)\}$
			\If{$TS_{\tau} \le s'_1$}  \Comment{below range}
			\State $b_{\tau} \gets b_1 + \dfrac{TS_{\tau}-s'_1}{s'_2 - s'_1}\,(b_2 - b_1)$
			\ElsIf{$TS_{\tau} \ge s'_n$}  \Comment{above range}
			\State $b_{\tau} \gets b_n + \dfrac{TS_{\tau}-s'_n}{s'_n - s'_{n-1}}\,(b_n - b_{n-1})$
			\Else  \Comment{within range}
			\State Find $i$ s.t.\ $s'_i \le TS_{\tau} \le s'_{i+1}$
			\State $b_{\tau} \gets b_i + \dfrac{TS_{\tau}-s'_i}{s'_{i+1} - s'_i}\,(b_{i+1} - b_i)$
			\EndIf
			\State \Return $b_{\tau}$
		\end{algorithmic}
	\end{algorithm}

	\subsection{Extension}
	We briefly discuss how to extend this framework to handle multi-constraint problems, such as \textit{cost cap bidding} for brand auction campaigns. The cost cap problem introduces an additional constraint on the average cost per conversion, increasing complexity by requiring the algorithm to balance both budget pacing and cost efficiency.
	
	The cost cap bidding problem can be formulated as an optimization problem that maximizes welfare utility subject to both a budget constraint and a cost per conversion constraint(e.g., see \cite{gao2022bidding}):
	\begin{equation} \label{eq:mpc_cost_cap}
		\begin{aligned}
			\max_{x_t \in \{0,1\}} \quad & \sum_{t=1}^{T} x_t \cdot r_t \\
			\text{s.t.} \quad & \sum_{t=1}^{T} x_t \cdot c_t \leq B, \\
			& \frac{\sum_{t=1}^{T} x_t \cdot c_t}{\sum_{t=1}^{T} x_t \cdot r_t} \leq C,
		\end{aligned}
	\end{equation}
	where \( C \) represents the cost per conversion cap specified by advertisers. The second constraint ensures that the average cost per conversion does not exceed the cap \( C \). The optimal bid per impression for problem~\eqref{eq:mpc_cost_cap} at time \( t \) is given by
	\[
	b^*_{t, \text{imp}} = \frac{1 + \mu^* C}{\lambda^* + \mu^*} \cdot r_t = b^* \cdot r_t,
	\]
	where \( \lambda^* \) and \( \mu^* \) are the Lagrangian dual variables corresponding to the budget and cost constraints, respectively, and \( b^* \) represents the optimal bid per conversion result for the campaign.
	
	At each time step \( \tau \), we solve the \textit{Receding Horizon Control (RHC)} problem to optimize the objective. The updated budget \( B_\tau \) is simply the remaining budget at time \( \tau \). The adjusted cost cap \( C_\tau \) ensures that the overall cost per conversion remains under the original cap \( C \), and is computed as
	\[
	C_\tau = \frac{B_\tau}{\frac{B}{C} - NC_\tau},
	\]
	where \( NC_\tau \) is the number of conversions observed up to time \( \tau \). The RHC optimization problem at time \( \tau \) is formulated as:
	\begin{equation} \label{eq:mpc_cost_cap_receding}
		\begin{aligned}
			\max_{x_t \in \{0,1\}} \quad & \sum_{t > \tau} x_t \cdot r_t \\
			\text{s.t.} \quad & \sum_{t > \tau} x_t \cdot c_t \leq B_\tau, \\
			& \frac{\sum_{t > \tau} x_t \cdot c_t}{\sum_{t > \tau} x_t \cdot r_t} \leq C_\tau.
		\end{aligned}
	\end{equation}
	
	The target spend \( TS_\tau \) for the next pacing interval \( (\tau, \tau + \Delta t) \) is computed as
	\[
	TS_\tau = B_\tau \cdot \frac{N_{\tau, \Delta t}}{T - \tau},
	\]
	and the target cost cap is set as 
	\[
	 TC_\tau = C_\tau.
	 \]
	
	The objective now is to determine the highest bid \( b \) such that the spend and cost per conversion over the next pacing interval do not exceed the constraints \( TS_\tau \) and \( TC_\tau \). For the budget constraint, we use the bid-to-spend model \( f(b) \), while for the cost constraint, we construct a bid-to-cost-per-result model \( h(b) \). To compute \( h(b) \), we collect the most recent \( N \) pacing intervals and construct \( \{ (b_k, n_k) \} \), where \( n_k \) is the number of conversions in interval \( k \). Applying the PAVA algorithm to \( \{ (b_k, n_k) \} \), we obtain a monotonic bid-to-conversions model \( n = g(b) \), and compute 
	
	\[
	h(b) = \frac{f(b)}{g(b)}.
	\]
	
	The final bid for the next pacing cycle is selected as the largest value \( b \) such that \( f(b) \leq TS_\tau \) and \( h(b) \leq TC_\tau \). This procedure is summarized in Algorithm \autoref{alg:mpc_cost_cap}.

	\begin{algorithm}[t]
		\caption{MPC-Based Cost Cap Bidding Algorithm}
		\label{alg:mpc_cost_cap}
		\begin{algorithmic}[1]
			\State \textbf{Input:} \\
			\hspace{1em}  $T$: Total predicted number of auction opportunities; $B$: Total budget; $C$: Cost cap; \\
			\hspace{1em} $\tau$: \(\tau\)-th auction request; $B_{\tau}$: Remaining budget; $NC_{\tau}$: Observed conversions;  \\
			\hspace{1em} $ b_u$: Max bid value; $\Delta b$: Search step size. 
			\State \textbf{Output:} $b^*$: Optimal bid for next pacing interval.
			
			\Statex
			\State \textbf{Step 1: Compute Budget and Cost Cap Constraints}
			\State Compute the target spend for next pacing interval:
			\[
			TS_{\tau} \gets  B_{\tau} \cdot \frac{N_{\tau, \Delta t}}{T - \tau}
			\]
			\State Compute the  cost cap \(TC_{\tau}\) for the next pacing interval: 
			\[
			TC_{\tau} \gets  \frac{B_{\tau}}{ \frac{B}{C} - NC_{\tau} }
			\]
			
			\Statex
			\State \textbf{Step 2: Construct Models $f(b)$ and $h(b)$}
			\State Use recent $N$ bid-spend pairs $\{(b_k, s_k)\}$ and apply \autoref{alg:pava} to build $f(b)$.
			\State Use recent $N$ bid-conversion pairs $\{(b_k, n_k)\}$ and apply  \autoref{alg:pava} to build $g(b)$.
			\State Compute cost per conversion estimate:
			\[
			h(b) \gets \frac{f(b)}{g(b)}
			\]
			
			\Statex
			\State \textbf{Step 3: Search for Optimal Bid $b^*$}
			\State Initialize $b^* \gets 0$.
			\For{$b \gets 0 $ to $b_u$ with step size $\Delta b$}
			\If{$f(b) \leq TS_{\tau}$ \textbf{and} $h(b) \leq TC_{\tau}$}
			\State Update $b^* \gets b$
			\EndIf
			\EndFor
			
			\Statex
			\State \textbf{Step 4: Return Optimal Bid}
			\State \Return $b^*$
			
		\end{algorithmic}
	\end{algorithm}

	\section{Evaluation}
	In this section, we present empirical results to demonstrate the effectiveness of the proposed MPC framework. We first conduct offline simulations to evaluate the performance of our MPC bidding strategy against several popular benchmarks, including PID control and gradient descent-based methods. We then present results from real-world online A/B testing on large-scale brand auction ads conducted at TikTok, further validating the practical effectiveness of our approach in production environments.

	\subsection{Offline Simulations}
	
	We conduct simulations to quantitatively evaluate the performance of our proposed bidding algorithms.
	
	\subsubsection{Benchmarks}  
	We compare our lightweight MPC against two industry-standard pacing strategies.
	
	\medskip
	
	\noindent\textbf{PID.} Following \cite{zhang2016feedback}, at each pacing interval starting at \(\tau\) we compute the normalized error  
	\[
	e_{\tau} = 1 - \frac{AS_{\tau}}{TS_{\tau}},
	\]
	where \(AS_{\tau}\) is the actual spend and \(TS_{\tau}\) is the target spend in that interval. The bid is then adjusted based on the PID controller update rule propsed in \cite{zhang2016feedback}.
	
	\medskip
	
	\noindent\textbf{Dual Online Gradient Descent (DOGD).}  Following \cite{gao2022bidding}, at each pacing interval \(\tau\) we update the dual variable \(\lambda\) via
	\[
	\lambda \leftarrow \lambda \;-\;\epsilon_{\tau} \cdot \,\biggl(1 \;-\;\frac{AS_{\tau}/NR_{\tau}}{B/T}\biggr),
	\]
	where \(AS_{\tau}\) is the actual spend observed in the pacing interval at \(\tau\), \(NR_{\tau}\) is the number of auction requests, \(B\) is the total budget, \(T\) is the total predicted number of auction opportunities, and \(\epsilon_{\tau}\) is the learning rate at \(\tau\). The bid per conversion is then set to \(1/\lambda\) according to \eqref{eq:optimal_bid}.

%
%

	\subsubsection{Experiment Set-up}
	We conduct simulations using a daily paced video ad campaign. The daily budget is denoted by \( B \), and for cost cap campaign simulations, the cost cap is set to \( C \). The total predicted number of auction opportunities is \( T \). The video play rate (VPR) \( r_t \) is sampled from a log-normal distribution \( \text{LogNorm}(\mu_1, \sigma_1) \), and the second price \( c_t \) is sampled from \( \text{LogNorm}(\mu_2, \sigma_2) \). The parameters \(\mu_1\), \(\sigma_1\), \(\mu_2\), and \(\sigma_2\) are estimated by fitting the corresponding distributions to internal auction data from TikTok advertising platform.
	
	For each bidding algorithm (MPC and benchmarks), each simulation episode continues until either the campaign budget is fully depleted or all auction opportunities are exhausted. We run 100 episodes for each algorithm and report the average of key evaluation metrics to assess performance. 

	\subsubsection{Result Analysis}
	Several evaluation metrics are computed to assess the effectiveness of the bidding strategies, including the \textit{budget utilization rate (BUR)} to measure pacing efficiency, the \textit{cost per video view (CPV)} to evaluate return on investment (ROI) for advertisers, and the \textit{bid variance (BV)} to assess the stability of pacing dynamics.
	
	The BUR is defined as the ratio of the delivered budget to the total allocated budget. To compute the cost per video view (CPV), we approximate the total number of video views by summing the sampled VPR values across all winning auction opportunities. The BV is defined as:
	\[
	BV = \frac{1}{T} \sum_{t=1}^{T} \left(1 - \frac{b_t}{b_m}\right)^2,
	\]
	where \( b_t \) is the bid at time \( t \), and \( b_m \) is the average bid across all bids during the pacing day. A lower BV indicates more stable bidding behavior, leading to smoother budget pacing and delivery. 
	
	We first perform a grid search to determine the optimal constant bid, which serves as the reference point for comparison. We then fine-tune the PID controller (by adjusting the control gains) and the DOGD algorithm (by tuning the learning rate) to identify their best-performing configurations. The simulation results are summarized in \autoref{tab:algo_comparison}. As shown in the results, all the fine-tuned algorithms successfully utilize the full budget. Among them, the MPC approach achieves the best ROI metrics, reflected by the lowest CPV. This improvement is partially attributed to its reduction of bid variance, which helps stabilize the bidding dynamics.

	\begin{table}[t]
		\caption{Performance Comparison of Max Delivery }
		\label{tab:algo_comparison}
		\centering
		\begin{tabular}{lcccc}
			\toprule
			\textbf{Algorithm} & \textbf{BUR (\%)} & \textbf{\#Impressions} & \textbf{CPV} & \textbf{BV} \\
			\midrule
			PID      & 99.9  & 23,759  & 0.01360   & 0.0860 \\
			DOGD     & 99.9  & 22,754    & 0.01422  & 0.1653 \\
			MPC      & 99.8  & 24,582    & 0.01319  & 0.0346 \\
			Optimal  & 99.9   &24,584   & 0.01318   & 0.00 \\
			\bottomrule
		\end{tabular}
	\end{table}
	
\begin{figure}[t]
		\centering
		\includegraphics[width=\linewidth]{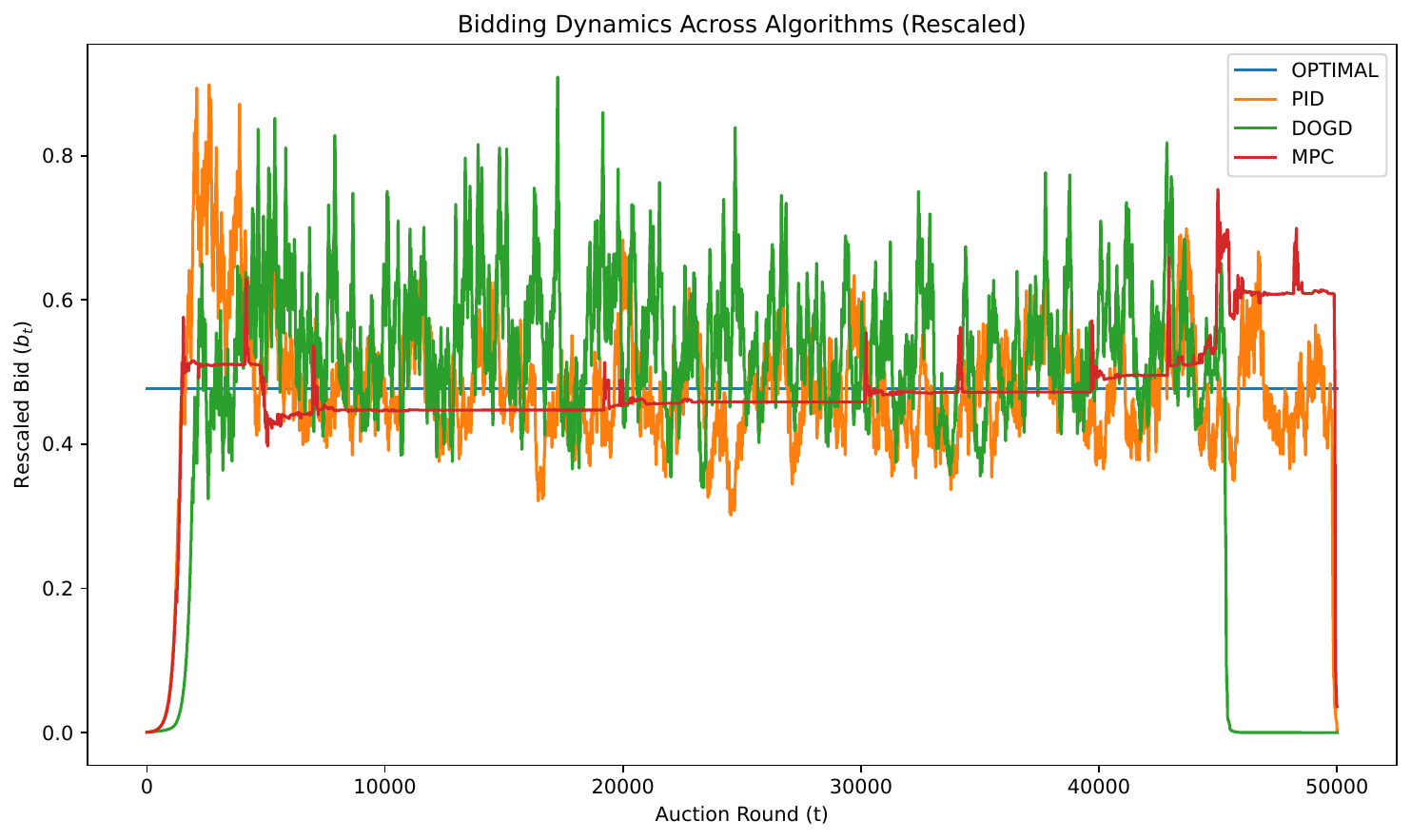}
		\caption{Sampled path of bid of the tested algorithms. Bid is rescaled to [0, 1]}
		\label{fig:bidding_dynamics}
	\end{figure}

	To further validate our hypothesis, we plot a sampled path of bid(rescaled to \([0, 1]\)) to illustrate the pacing dynamics of each algorithm in \autoref{fig:bidding_dynamics}. The MPC algorithm exhibits significantly more stable and controlled bidding behavior compared to PID and DOGD. We also observe that both PID and DOGD tend to deplete the budget before the end of the pacing day, potentially due to their lack of planning capability, which MPC effectively addresses through its receding horizon optimization.
	
\begin{figure}[t]
		\centering
		\includegraphics[width=\linewidth]{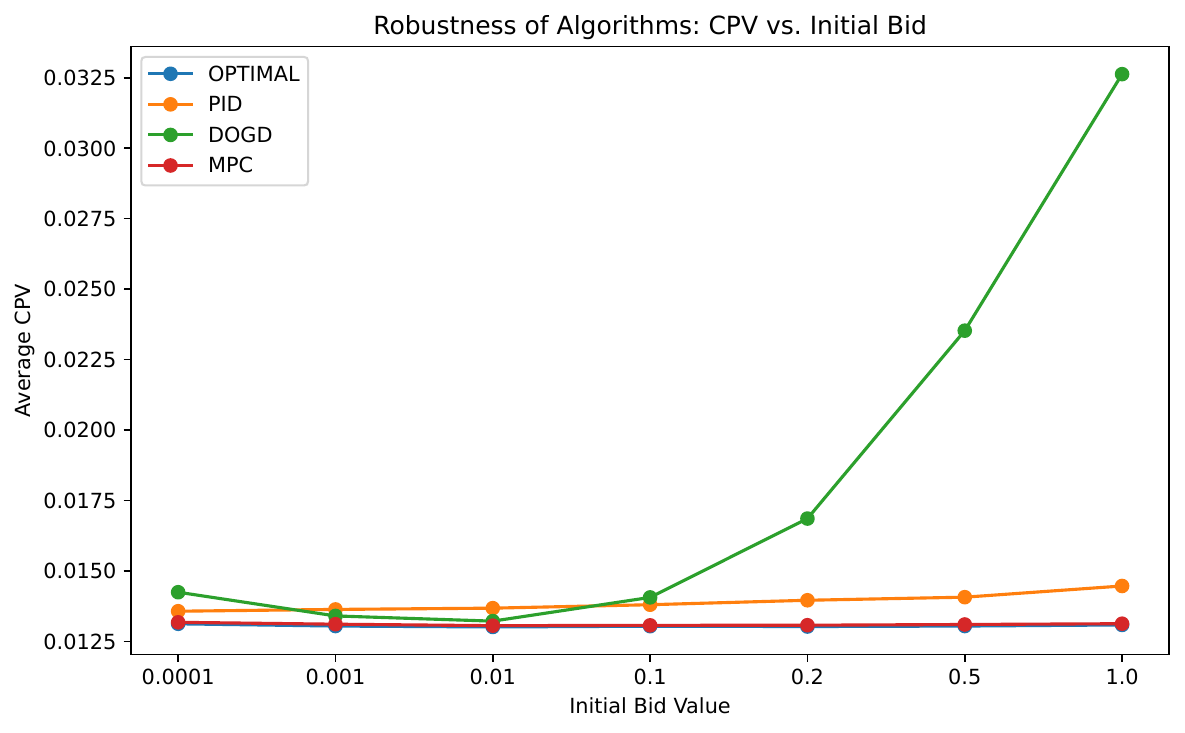}
		\caption{CPV versus initial bid value for each algorithm. MPC is highly robust to initialization, while PID and DOGD suffer as the cold-start bid deviates from the true optimum.}
		\label{fig:init_value}
	\end{figure}

	In practice, the initial bid for a cold start campaign is estimated offline, and a bidding strategy's performance is sometimes highly sensitive to this estimate. To evaluate robustness to this factor, we vary the initial bid across seven values from 0.0001 to 1.0 and measure the resulting metrics. We observe that all algorithms fully deplete their budget in each run. To compare the ROI, average CPV versus initial bid is plotted in Figure \ref{fig:init_value}: MPC's performance remains essentially invariant and closely tracks the optimal ROI, while the other two---and in particular DOGD---exhibit significant performance degradation as the initialization error grows.

	\subsection{Large Scale Online A/B Test}
	
	We conduct a budget split experiment on the TikTok advertising platform to evaluate the proposed MPC framework. The experiment involves tens of thousands of brand auction campaigns, including max delivery and cost cap, with both awareness and video view objectives, over a 7-day period. The MPC framework is tested against the existing production bidding framework. Experimental results show significant improvements across multiple key performance metrics, including budget utilization rate, CPM, and CPV.

	\section{Remarks}
	The unique attributes of brand auction ads make this lightweight framework particularly effective. The charging model we use in this paper is on a per-impression basis. In practice, the charging models for brand ads can vary; for example, video ads typically are charged for different objectives such as 2-second, 5-second, or 15-second views. This introduces delays in the feedback loop; however, in online experiments, we find that such impact is relatively small on performance, even for 15-second video ads.
	
	In contrast, performance ads typically involve deeper funnel engagement events and are often charged based on specific user actions (e.g., clicks, conversions). This results in much sparser spend signals and makes it challenging to directly extend our online MPC framework to such scenarios. Even when using per-impression charging to mitigate spend signal sparsity, the elongated conversion funnels associated with performance ads introduce higher variance in spend data. To see this, recall that to construct the bid-to-spend-rate model \( f \), we collect recent bid-spend pairs from each pacing interval. The spend signal aggregates costs across all auction opportunities within the interval, where each auction request may have a different conversion rate \( r_t \). Our framework relies on the assumption that these conversion rates \( r_t \) are random and approximately i.i.d., allowing statistical aggregation to reduce variance and produce robust bid-spend modeling. This assumption holds reasonably well for upper-funnel objectives such as video views, where the conversion funnel is short and feedback is timely. However, for performance ads involving post-click deep-funnel actions, outcomes become significantly more volatile.
	
	Deep-funnel cost control further compounds these challenges, introducing both signal sparsity and highly noisy conversion estimates. In such cases, more sophisticated techniques---such as model calibration, delayed feedback handling, and advanced variance reduction methods---should be incorporated to make the online MPC framework effective and reliable.

	\section{Conclusion}
	This work presents a practical solution for optimizing brand advertising campaigns through a lightweight MPC-based bidding framework. By fully exploiting the predictable impression patterns and timely feedback loops inherent in brand auction ads, our framework simplifies bid decision-making while effectively balancing multiple business objectives, including budget pacing and cost efficiency. Unlike complex machine learning-based approaches, our method relies on simple, interpretable models that can be deployed entirely online with minimal system overhead. Both offline simulations and large-scale A/B testing on the TikTok platform validate the effectiveness of our approach in real-world environments. While the current framework is well-suited for upper-funnel brand objectives, extending it to handle the delayed and sparse feedback typical of performance advertising remains an important direction for future research.


	\appendix
	
	\section{Proof of Optimal Bidding Bidding Formula}  
	\label{appendix:optimal_bidding}
	For completeness, we include below the proof of the optimal-bidding formula \eqref{eq:optimal_bid} for the max-delivery problem \eqref{eq:max_delivery}. Under the regularity assumption that \(\{r_t\}\) and \(\{c_t\}\) are sampled independently from two different i.i.d.\ distributions, one can show that the optimal bid level causes actual spend to scale linearly with the number of auction opportunities. Similar derivations can be found in \cite{balseiro2019learning, fernandez2017optimal, gao2022bidding}.
	
	\begin{proof}
	We start from the primal max delivery problem
	\[
	\max_{x_t\in\{0,1\}}\;\sum_{t=1}^T x_t\,r_t
	\quad\text{s.t.}\quad
	\sum_{t=1}^T x_t\,c_t \le B.
	\]
	Introducing the dual variable \(\lambda\ge0\), the Lagrangian is
	\[
	\mathcal{L}(x,\lambda)
	= \sum_{t=1}^T x_t\,r_t \;-\;\lambda\Bigl(\sum_{t=1}^T x_t\,c_t - B\Bigr)
	= \sum_{t=1}^T x_t\,(r_t - \lambda\,c_t) \;+\;\lambda\,B.
	\]
	Thus the dual function is
	\[
	\mathcal{L}^*(\lambda)
	= \max_{x_t\in\{0,1\}}\mathcal{L}(x,\lambda)
	= \sum_{t=1}^T \bigl(r_t - \lambda\,c_t\bigr)_+ \;+\;\lambda\,B,
	\]
	where \((z)_+=\max\{z,0\}\).  The dual problem is then
	\[
	\min_{\lambda\ge0}\;\mathcal{L}^*(\lambda)
	= \min_{\lambda\ge0}\;\sum_{t=1}^T \Bigl[(r_t - \lambda\,c_t)_+ + \lambda\,\tfrac{B}{T}\Bigr].
	\]
	Let \(\lambda^*\) be its minimizer.  By KKT, if \(\lambda^*>0\) then the budget constraint binds:
	\[
	\sum_{t=1}^T x_t\,c_t = B.
	\]
	Moreover, the optimal binary decision is
	\[
	x_t^* = 
	\begin{cases}
		1, & r_t - \lambda^*\,c_t > 0,\\
		0, & \text{otherwise},
	\end{cases}
	\]
	so the corresponding optimal bid is 
	\[
	b_t^* \;=\;\frac{r_t}{\lambda^*}.
	\]
	Under an i.i.d. model for \(r_t\) and \(c_t\), each served auction has constant expected cost, and hence over any window \(\Delta t\),
	\[
	\sum_{\tau\le t\le\tau+\Delta t} x_t\,c_t 
	\;\propto\;
	\#\{\text{auctions in }(\tau,\tau+\Delta t)\}.
	\]
	\end{proof}	


\begin{thebibliography}{31}


\ifx \showCODEN    \undefined \def \showCODEN     #1{\unskip}     \fi
\ifx \showISBNx    \undefined \def \showISBNx     #1{\unskip}     \fi
\ifx \showISBNxiii \undefined \def \showISBNxiii  #1{\unskip}     \fi
\ifx \showISSN     \undefined \def \showISSN      #1{\unskip}     \fi
\ifx \showLCCN     \undefined \def \showLCCN      #1{\unskip}     \fi
\ifx \shownote     \undefined \def \shownote      #1{#1}          \fi
\ifx \showarticletitle \undefined \def \showarticletitle #1{#1}   \fi
\ifx \showURL      \undefined \def \showURL       {\relax}        \fi
\providecommand\bibfield[2]{#2}
\providecommand\bibinfo[2]{#2}
\providecommand\natexlab[1]{#1}
\providecommand\showeprint[2][]{arXiv:#2}

\bibitem[Abrams et~al\mbox{.}(2008)]%
        {abrams2008ad}
\bibfield{author}{\bibinfo{person}{Zo{\"o} Abrams}, \bibinfo{person}{S~Sathiya
  Keerthi}, \bibinfo{person}{Ofer Mendelevitch}, {and} \bibinfo{person}{John~A
  Tomlin}.} \bibinfo{year}{2008}\natexlab{}.
\newblock \showarticletitle{Ad delivery with budgeted advertisers: A
  comprehensive LP approach.}
\newblock \bibinfo{journal}{\emph{Journal of Electronic Commerce Research}}
  \bibinfo{volume}{9}, \bibinfo{number}{1} (\bibinfo{year}{2008}).
\newblock


\bibitem[Agarwal et~al\mbox{.}(2014)]%
        {agarwal2014budget}
\bibfield{author}{\bibinfo{person}{Deepak Agarwal}, \bibinfo{person}{Souvik
  Ghosh}, \bibinfo{person}{Kai Wei}, {and} \bibinfo{person}{Siyu You}.}
  \bibinfo{year}{2014}\natexlab{}.
\newblock \showarticletitle{Budget pacing for targeted online advertisements at
  linkedin}. In \bibinfo{booktitle}{\emph{Proceedings of the 20th ACM SIGKDD
  international conference on Knowledge discovery and data mining}}.
  \bibinfo{pages}{1613--1619}.
\newblock


\bibitem[Aggarwal et~al\mbox{.}(2019)]%
        {aggarwal2019autobidding}
\bibfield{author}{\bibinfo{person}{Gagan Aggarwal},
  \bibinfo{person}{Ashwinkumar Badanidiyuru}, {and} \bibinfo{person}{Aranyak
  Mehta}.} \bibinfo{year}{2019}\natexlab{}.
\newblock \showarticletitle{Autobidding with constraints}. In
  \bibinfo{booktitle}{\emph{Web and Internet Economics: 15th International
  Conference, WINE 2019, New York, NY, USA, December 10--12, 2019, Proceedings
  15}}. Springer, \bibinfo{pages}{17--30}.
\newblock


\bibitem[Ayer et~al\mbox{.}(1955)]%
        {ayer1955empirical}
\bibfield{author}{\bibinfo{person}{Miriam Ayer}, \bibinfo{person}{H~Daniel
  Brunk}, \bibinfo{person}{George~M Ewing}, \bibinfo{person}{William~T Reid},
  {and} \bibinfo{person}{Edward Silverman}.} \bibinfo{year}{1955}\natexlab{}.
\newblock \showarticletitle{An empirical distribution function for sampling
  with incomplete information}.
\newblock \bibinfo{journal}{\emph{The annals of mathematical statistics}}
  (\bibinfo{year}{1955}), \bibinfo{pages}{641--647}.
\newblock


\bibitem[Balseiro et~al\mbox{.}(2017)]%
        {balseiro2017budget}
\bibfield{author}{\bibinfo{person}{Santiago Balseiro}, \bibinfo{person}{Anthony
  Kim}, \bibinfo{person}{Mohammad Mahdian}, {and} \bibinfo{person}{Vahab
  Mirrokni}.} \bibinfo{year}{2017}\natexlab{}.
\newblock \showarticletitle{Budget management strategies in repeated auctions}.
  In \bibinfo{booktitle}{\emph{Proceedings of the 26th International Conference
  on World Wide Web}}. \bibinfo{pages}{15--23}.
\newblock


\bibitem[Balseiro and Gur(2019)]%
        {balseiro2019learning}
\bibfield{author}{\bibinfo{person}{Santiago~R Balseiro} {and}
  \bibinfo{person}{Yonatan Gur}.} \bibinfo{year}{2019}\natexlab{}.
\newblock \showarticletitle{Learning in repeated auctions with budgets: Regret
  minimization and equilibrium}.
\newblock \bibinfo{journal}{\emph{Management Science}} \bibinfo{volume}{65},
  \bibinfo{number}{9} (\bibinfo{year}{2019}), \bibinfo{pages}{3952--3968}.
\newblock


\bibitem[Balseiro et~al\mbox{.}(2022)]%
        {balseiro2022analysis}
\bibfield{author}{\bibinfo{person}{Santiago~R Balseiro},
  \bibinfo{person}{Haihao Lu}, \bibinfo{person}{Vahab Mirrokni}, {and}
  \bibinfo{person}{Balasubramanian Sivan}.} \bibinfo{year}{2022}\natexlab{}.
\newblock \showarticletitle{Analysis of dual-based pid controllers through
  convolutional mirror descent}.
\newblock \bibinfo{journal}{\emph{arXiv preprint arXiv:2202.06152}}
  (\bibinfo{year}{2022}).
\newblock


\bibitem[Barlow(1972)]%
        {barlow1972statistical}
\bibfield{author}{\bibinfo{person}{Richard~E Barlow}.}
  \bibinfo{year}{1972}\natexlab{}.
\newblock \showarticletitle{Statistical inference under order restrictions: The
  theory and application of isotonic regression}.
\newblock \bibinfo{journal}{\emph{(No Title)}} (\bibinfo{year}{1972}).
\newblock


\bibitem[Busing(2022)]%
        {busing2022monotone}
\bibfield{author}{\bibinfo{person}{Frank~MTA Busing}.}
  \bibinfo{year}{2022}\natexlab{}.
\newblock \showarticletitle{Monotone regression: A simple and fast O (n) PAVA
  implementation}.
\newblock \bibinfo{journal}{\emph{Journal of Statistical Software}}
  \bibinfo{volume}{102} (\bibinfo{year}{2022}), \bibinfo{pages}{1--25}.
\newblock


\bibitem[Castiglioni et~al\mbox{.}(2022)]%
        {castiglioni2022online}
\bibfield{author}{\bibinfo{person}{Matteo Castiglioni}, \bibinfo{person}{Andrea
  Celli}, {and} \bibinfo{person}{Christian Kroer}.}
  \bibinfo{year}{2022}\natexlab{}.
\newblock \showarticletitle{Online learning with knapsacks: the best of both
  worlds}. In \bibinfo{booktitle}{\emph{International Conference on Machine
  Learning}}. PMLR, \bibinfo{pages}{2767--2783}.
\newblock


\bibitem[Chen et~al\mbox{.}(2014)]%
        {chen2014dynamic}
\bibfield{author}{\bibinfo{person}{Bowei Chen}, \bibinfo{person}{Shuai Yuan},
  {and} \bibinfo{person}{Jun Wang}.} \bibinfo{year}{2014}\natexlab{}.
\newblock \showarticletitle{A dynamic pricing model for unifying programmatic
  guarantee and real-time bidding in display advertising}. In
  \bibinfo{booktitle}{\emph{Proceedings of the Eighth International Workshop on
  Data Mining for Online Advertising}}. \bibinfo{pages}{1--9}.
\newblock


\bibitem[Chen(2025)]%
        {chen2025practical}
\bibfield{author}{\bibinfo{person}{Yuanlong Chen}.}
  \bibinfo{year}{2025}\natexlab{}.
\newblock \showarticletitle{A Practical Guide to Budget Pacing Algorithms in
  Digital Advertising}.
\newblock \bibinfo{journal}{\emph{arXiv preprint arXiv:2503.06942}}
  (\bibinfo{year}{2025}).
\newblock


\bibitem[Chen et~al\mbox{.}(2011)]%
        {chen2011real}
\bibfield{author}{\bibinfo{person}{Ye Chen}, \bibinfo{person}{Pavel Berkhin},
  \bibinfo{person}{Bo Anderson}, {and} \bibinfo{person}{Nikhil~R Devanur}.}
  \bibinfo{year}{2011}\natexlab{}.
\newblock \showarticletitle{Real-time bidding algorithms for performance-based
  display ad allocation}. In \bibinfo{booktitle}{\emph{Proceedings of the 17th
  ACM SIGKDD international conference on Knowledge discovery and data mining}}.
  \bibinfo{pages}{1307--1315}.
\newblock


\bibitem[Conitzer et~al\mbox{.}(2022)]%
        {conitzer2022multiplicative}
\bibfield{author}{\bibinfo{person}{Vincent Conitzer},
  \bibinfo{person}{Christian Kroer}, \bibinfo{person}{Eric Sodomka}, {and}
  \bibinfo{person}{Nicolas~E Stier-Moses}.} \bibinfo{year}{2022}\natexlab{}.
\newblock \showarticletitle{Multiplicative pacing equilibria in auction
  markets}.
\newblock \bibinfo{journal}{\emph{Operations Research}} \bibinfo{volume}{70},
  \bibinfo{number}{2} (\bibinfo{year}{2022}), \bibinfo{pages}{963--989}.
\newblock


\bibitem[Dai et~al\mbox{.}(2024)]%
        {dai2024percentile}
\bibfield{author}{\bibinfo{person}{Liang Dai}, \bibinfo{person}{Kejie Lyu},
  \bibinfo{person}{Chengcheng Zhang}, \bibinfo{person}{Guangming Zhao},
  \bibinfo{person}{Zhonglin Zu}, \bibinfo{person}{Liang Wang}, {and}
  \bibinfo{person}{Bo Zheng}.} \bibinfo{year}{2024}\natexlab{}.
\newblock \showarticletitle{Percentile risk-constrained budget pacing for
  guaranteed display advertising in online optimization}. In
  \bibinfo{booktitle}{\emph{Proceedings of the AAAI Conference on Artificial
  Intelligence}}, Vol.~\bibinfo{volume}{38}. \bibinfo{pages}{7987--7994}.
\newblock


\bibitem[Fang et~al\mbox{.}(2019)]%
        {fang2019large}
\bibfield{author}{\bibinfo{person}{Zhen Fang}, \bibinfo{person}{Yang Li},
  \bibinfo{person}{Chuanren Liu}, \bibinfo{person}{Wenxiang Zhu},
  \bibinfo{person}{Yu Zheng}, {and} \bibinfo{person}{Wenjun Zhou}.}
  \bibinfo{year}{2019}\natexlab{}.
\newblock \showarticletitle{Large-scale personalized delivery for guaranteed
  display advertising with real-time pacing}. In \bibinfo{booktitle}{\emph{2019
  IEEE International Conference on Data Mining (ICDM)}}. IEEE,
  \bibinfo{pages}{190--199}.
\newblock


\bibitem[Feldman et~al\mbox{.}(2010)]%
        {feldman2010online}
\bibfield{author}{\bibinfo{person}{Jon Feldman}, \bibinfo{person}{Monika
  Henzinger}, \bibinfo{person}{Nitish Korula}, \bibinfo{person}{Vahab~S
  Mirrokni}, {and} \bibinfo{person}{Cliff Stein}.}
  \bibinfo{year}{2010}\natexlab{}.
\newblock \showarticletitle{Online stochastic packing applied to display ad
  allocation}. In \bibinfo{booktitle}{\emph{European Symposium on Algorithms}}.
  Springer, \bibinfo{pages}{182--194}.
\newblock


\bibitem[Fernandez-Tapia et~al\mbox{.}(2017)]%
        {fernandez2017optimal}
\bibfield{author}{\bibinfo{person}{Joaquin Fernandez-Tapia},
  \bibinfo{person}{Olivier Gu{\'e}ant}, {and} \bibinfo{person}{Jean-Michel
  Lasry}.} \bibinfo{year}{2017}\natexlab{}.
\newblock \showarticletitle{Optimal real-time bidding strategies}.
\newblock \bibinfo{journal}{\emph{Applied mathematics research express}}
  \bibinfo{volume}{2017}, \bibinfo{number}{1} (\bibinfo{year}{2017}),
  \bibinfo{pages}{142--183}.
\newblock


\bibitem[Gao et~al\mbox{.}(2022)]%
        {gao2022bidding}
\bibfield{author}{\bibinfo{person}{Yuan Gao}, \bibinfo{person}{Kaiyu Yang},
  \bibinfo{person}{Yuanlong Chen}, \bibinfo{person}{Min Liu}, {and}
  \bibinfo{person}{Noureddine~El Karoui}.} \bibinfo{year}{2022}\natexlab{}.
\newblock \showarticletitle{Bidding agent design in the linkedin ad
  marketplace}.
\newblock \bibinfo{journal}{\emph{arXiv preprint arXiv:2202.12472}}
  (\bibinfo{year}{2022}).
\newblock


\bibitem[Geyik et~al\mbox{.}(2020)]%
        {geyik2020impression}
\bibfield{author}{\bibinfo{person}{Sahin~Cem Geyik}, \bibinfo{person}{Luthfur
  Chowdhury}, \bibinfo{person}{Florian Raudies}, \bibinfo{person}{Wen Pu},
  {and} \bibinfo{person}{Jianqiang Shen}.} \bibinfo{year}{2020}\natexlab{}.
\newblock \showarticletitle{Impression Pacing for Jobs Marketplace at
  LinkedIn}. In \bibinfo{booktitle}{\emph{Proceedings of the 29th ACM
  International Conference on Information \& Knowledge Management}}.
  \bibinfo{pages}{2445--2452}.
\newblock


\bibitem[Grigas et~al\mbox{.}(2017)]%
        {grigas2017profit}
\bibfield{author}{\bibinfo{person}{Paul Grigas}, \bibinfo{person}{Alfonso
  Lobos}, \bibinfo{person}{Zheng Wen}, {and} \bibinfo{person}{Kuang-chih Lee}.}
  \bibinfo{year}{2017}\natexlab{}.
\newblock \showarticletitle{Profit maximization for online advertising
  demand-side platforms}.
\newblock In \bibinfo{booktitle}{\emph{Proceedings of the ADKDD'17}}.
  \bibinfo{pages}{1--7}.
\newblock


\bibitem[Maehara et~al\mbox{.}(2018)]%
        {maehara2018optimal}
\bibfield{author}{\bibinfo{person}{Takanori Maehara}, \bibinfo{person}{Atsuhiro
  Narita}, \bibinfo{person}{Jun Baba}, {and} \bibinfo{person}{Takayuki
  Kawabata}.} \bibinfo{year}{2018}\natexlab{}.
\newblock \showarticletitle{Optimal Bidding Strategy for Brand Advertising.}.
  In \bibinfo{booktitle}{\emph{IJCAI}}. \bibinfo{pages}{424--432}.
\newblock


\bibitem[Mehta et~al\mbox{.}(2007)]%
        {mehta2007adwords}
\bibfield{author}{\bibinfo{person}{Aranyak Mehta}, \bibinfo{person}{Amin
  Saberi}, \bibinfo{person}{Umesh Vazirani}, {and} \bibinfo{person}{Vijay
  Vazirani}.} \bibinfo{year}{2007}\natexlab{}.
\newblock \showarticletitle{Adwords and generalized online matching}.
\newblock \bibinfo{journal}{\emph{Journal of the ACM (JACM)}}
  \bibinfo{volume}{54}, \bibinfo{number}{5} (\bibinfo{year}{2007}),
  \bibinfo{pages}{22--es}.
\newblock


\bibitem[Nuara et~al\mbox{.}(2022)]%
        {nuara2022online}
\bibfield{author}{\bibinfo{person}{Alessandro Nuara},
  \bibinfo{person}{Francesco Trov{\`o}}, \bibinfo{person}{Nicola Gatti}, {and}
  \bibinfo{person}{Marcello Restelli}.} \bibinfo{year}{2022}\natexlab{}.
\newblock \showarticletitle{Online joint bid/daily budget optimization of
  internet advertising campaigns}.
\newblock \bibinfo{journal}{\emph{Artificial Intelligence}}
  \bibinfo{volume}{305} (\bibinfo{year}{2022}), \bibinfo{pages}{103663}.
\newblock


\bibitem[Tashman et~al\mbox{.}(2020)]%
        {tashman2020dynamic}
\bibfield{author}{\bibinfo{person}{Michael Tashman}, \bibinfo{person}{Jiayi
  Xie}, \bibinfo{person}{John Hoffman}, \bibinfo{person}{Lee Winikor}, {and}
  \bibinfo{person}{Rouzbeh Gerami}.} \bibinfo{year}{2020}\natexlab{}.
\newblock \showarticletitle{Dynamic bidding strategies with multivariate
  feedback control for multiple goals in display advertising}.
\newblock \bibinfo{journal}{\emph{arXiv preprint arXiv:2007.00426}}
  (\bibinfo{year}{2020}).
\newblock


\bibitem[Wang et~al\mbox{.}(2017)]%
        {wang2017display}
\bibfield{author}{\bibinfo{person}{Jun Wang}, \bibinfo{person}{Weinan Zhang},
  \bibinfo{person}{Shuai Yuan}, {et~al\mbox{.}}}
  \bibinfo{year}{2017}\natexlab{}.
\newblock \showarticletitle{Display advertising with real-time bidding (RTB)
  and behavioural targeting}.
\newblock \bibinfo{journal}{\emph{Foundations and Trends{\textregistered} in
  Information Retrieval}} \bibinfo{volume}{11}, \bibinfo{number}{4-5}
  (\bibinfo{year}{2017}), \bibinfo{pages}{297--435}.
\newblock


\bibitem[Wang et~al\mbox{.}(2022)]%
        {wang2022conflux}
\bibfield{author}{\bibinfo{person}{XiaoYu Wang}, \bibinfo{person}{Bin Tan},
  \bibinfo{person}{Yonghui Guo}, \bibinfo{person}{Tao Yang},
  \bibinfo{person}{Dongbo Huang}, \bibinfo{person}{Lan Xu},
  \bibinfo{person}{Nikolaos~M Freris}, \bibinfo{person}{Hao Zhou}, {and}
  \bibinfo{person}{Xiang-Yang Li}.} \bibinfo{year}{2022}\natexlab{}.
\newblock \showarticletitle{CONFLUX: A Request-level Fusion Framework for
  Impression Allocation via Cascade Distillation}. In
  \bibinfo{booktitle}{\emph{Proceedings of the 28th ACM SIGKDD Conference on
  Knowledge Discovery and Data Mining}}. \bibinfo{pages}{4070--4078}.
\newblock


\bibitem[Wu et~al\mbox{.}(2021)]%
        {wu2021impression}
\bibfield{author}{\bibinfo{person}{Di Wu}, \bibinfo{person}{Cheng Chen},
  \bibinfo{person}{Xiujun Chen}, \bibinfo{person}{Junwei Pan},
  \bibinfo{person}{Xun Yang}, \bibinfo{person}{Qing Tan}, \bibinfo{person}{Jian
  Xu}, {and} \bibinfo{person}{Kuang-Chih Lee}.}
  \bibinfo{year}{2021}\natexlab{}.
\newblock \showarticletitle{Impression Allocation and Policy Search in Display
  Advertising}. In \bibinfo{booktitle}{\emph{2021 IEEE International Conference
  on Data Mining (ICDM)}}. IEEE, \bibinfo{pages}{749--756}.
\newblock


\bibitem[Xu et~al\mbox{.}(2015)]%
        {xu2015smart}
\bibfield{author}{\bibinfo{person}{Jian Xu}, \bibinfo{person}{Kuang-chih Lee},
  \bibinfo{person}{Wentong Li}, \bibinfo{person}{Hang Qi}, {and}
  \bibinfo{person}{Quan Lu}.} \bibinfo{year}{2015}\natexlab{}.
\newblock \showarticletitle{Smart pacing for effective online ad campaign
  optimization}. In \bibinfo{booktitle}{\emph{Proceedings of the 21th ACM
  SIGKDD international conference on knowledge discovery and data mining}}.
  \bibinfo{pages}{2217--2226}.
\newblock


\bibitem[Yang et~al\mbox{.}(2019)]%
        {yang2019bid}
\bibfield{author}{\bibinfo{person}{Xun Yang}, \bibinfo{person}{Yasong Li},
  \bibinfo{person}{Hao Wang}, \bibinfo{person}{Di Wu}, \bibinfo{person}{Qing
  Tan}, \bibinfo{person}{Jian Xu}, {and} \bibinfo{person}{Kun Gai}.}
  \bibinfo{year}{2019}\natexlab{}.
\newblock \showarticletitle{Bid optimization by multivariable control in
  display advertising}. In \bibinfo{booktitle}{\emph{Proceedings of the 25th
  ACM SIGKDD international conference on knowledge discovery \& data mining}}.
  \bibinfo{pages}{1966--1974}.
\newblock


\bibitem[Zhang et~al\mbox{.}(2016)]%
        {zhang2016feedback}
\bibfield{author}{\bibinfo{person}{Weinan Zhang}, \bibinfo{person}{Yifei Rong},
  \bibinfo{person}{Jun Wang}, \bibinfo{person}{Tianchi Zhu}, {and}
  \bibinfo{person}{Xiaofan Wang}.} \bibinfo{year}{2016}\natexlab{}.
\newblock \showarticletitle{Feedback control of real-time display advertising}.
  In \bibinfo{booktitle}{\emph{Proceedings of the Ninth ACM International
  Conference on Web Search and Data Mining}}. \bibinfo{pages}{407--416}.
\newblock


\end{thebibliography}
\end{document}